# Quality thinning and value development of boreal trees

Petri P. Kärenlampi[*]

Lehtoi Research, Finland

petri.karenlampi@professori.fi

[*] Author to whom correspondence should be addressed.

## Abstract

For the first time, quality distribution of trees is introduced in a tree growth model. Consequently, the effects of quality thinning on stand development can be investigated. Quality thinning improves the financial return in all cases studied, but the effect is small. Rotation ages, timber stocks and maturity diameters are not much affected by quality thinning. Bare land valuation neither changes the contribution of the quality thinning. The reason for the small effect apparently lies in the value development of individual trees. The relative value development of small pulpwood trunks is large, since the harvesting expense per volume unit is reduced along with size increment. Such trees are not feasible objects for quality thinning, unless quality correlates with growth rate. Another enhanced stage of value development is when pulpwood trunks turn to sawlog trunks. For large pulpwood trunks, quality thinning is feasible. Existing sawlog content in trees dilutes the effect of quality thinning on the financial return. The results change if the growth rate is positively correlated with quality, quality thinning becoming feasible in all commercial diameter classes.

## Introduction

As trees are individuals, their productive capacity, as well as their quality charateristics vary. Many forestry practices contain an idea of improvement of the quality distribution of remaining trees through quality thinning, or "selective thinning" [Phillips 2024, Nuutinen et al. 2021, 2023, Niemistö et al. 2018, Mäkinen et al. 2006, Karlsson et al. 2012, Segtowich et al. 2023, Cameron 2002]. However, there is no established definition of such a procedure, neither is the concept "quality" unambiguously defined. Some authors refer as quality thinning to a process which does not prioritize large or small trees, but rather retains trees of good quality in all size classes [Niemistö et al. 2018].

It has been stated that selective thinnings may improve resilience against snow and wind damage [Cameron 2002, Cremer et al. 1982, Persson 1972, Valinger et al. 1993]. It is apparently unknow how the selection process might differently contribute to resilience on the one hand, and the quality and value yield on the other.

It has been recognized that single "average" values of input parameters do not necessarily suffice as input for growth and yield models [Green and Strawderman 1996, MacFarlane et al. 2000]. However, connection of quality distributions of trees to operative stand management appears to be missing in the literature. This paper attempts to establish such a connection.

A problem in the establishment of the connection of quality distribution to stand management is the lack of an unambiguous definition of quality. Such a definition missing, the corresponding property can neither be measured objectively. A consequence is that the harvester operator must evaluate quality based on individual visual assessment, possibly assisted by instructions given by stakeholders. Then, the conception of quality will necessarily vary between operators.

The concept of quality obviously differs not only between operators, but between tree species, branches of utilization, as well as between geographic areas and operating companies. Some industries value low branchiness and knottiness, while others find trunk straightness more important. It is not uncommon to understand visually observable vigor as a quality characteristic.

As there is no universally valid measurable meaning of quality, instead of direct measurement, another kind of approach is needed in the investigation of quality thinning. Here, quality becomes defined through its effect on product value: high-quality trees produce more valuable products. Such a definition may appear awkward, but we will soon find that it is not difficult to quantify.

A basic assumption here is that the harvester operator is, at will, able to visually determine quality. Secondly, visually observable vigor may or may not be included in the concept of quality. In the positive (negative) case, quality affects product value, and (but not) the growth rate of trees.

In the remaining part of this paper, a distribution of the quality and the corresponding value of trees within any species and diameter class is introduced. The quality distribution is incorporated within an inventory-based growth model, and applied to a dataset of never-thinned, spruce-dominated boreal stands. Thinning procedures are designed for the stands with and without quality thinning, with the objective of maximizing the expected value of the operative return rate on capital. Finally, a correlation between quality and growth rate is established, and the effect of quality thinning is such a case is examined.

**Methods and Materials**

*Applied growth model*

An inventory-based growth model [Bollandsås et al. 2008] is applied to initial conditions resulting from field measurements of seven never-thinned forest stands. For any tree species and diameter class, the expected diameter change rate is computed for 30-month periods.

*Applied financial model*

We apply a procedure first mentioned in the literature in 1967, but applied only recently [Speidel 1967 1972,Kärenlampi 2019a, 2020a, 2020b, 2021a, 2021b, 2022a]. Instead of discounting revenues, the return rate on capital achieved as relative value increment at different stages of forest stand development is weighed by current capitalization, and integrated.

The return rate on capital is the relative time change rate of value. We choose to write

$$r(t) = \frac{d\kappa}{K(t)dt}$$ (1)

where $\kappa$ in the numerator considers value growth, operative expenses, and amortizations, but neglects investments and withdrawals. In other words, it is the change of capitalization on an operating profit basis. $K$ in the denominator gives capitalization on a balance sheet basis, being directly affected by any investment or withdrawal.

A momentary rate of the rate of return on capital of Eq. (1) naturally is not representative for an entire forestry system – an expected value of the return rate is needed for management considerations. As time proceeds linearly, the expected value of the return rate on capital within the rotation can be written as [Kärenlampi 2022b, 2022a]

$$\langle r(t) \rangle = \frac{\int_{b}^{b+\tau} \frac{d\kappa}{dt}(a)da}{\int_{b}^{b+\tau} K(a)da} = \frac{\int_{b}^{b+\tau} r(a)K(a)da}{\int_{b}^{b+\tau} K(a)da}$$ (2),

where $a$ is stand age, $b$ is arbitrary starting age of the integration, and $\tau$ is rotation age. Along with a periodic boundary condition, Eq. (2) does not depend on the starting point of the integration.

*Quality distributions and quality thinning*

Instead of having a constant value per trunk or per cubic meter within a class of trees of specified tree species and size, the properties vary. The expected monetary value of the assortments from any tree is determined based on the expected yield of pulpwood and sawlogs, along with the prices of the assortments [Kärenlampi 2022d]. Here, the distribution of quality is modeled as a uniform distribution between a minimum and a maximum, these being placed symmetrically with respect to the mean. Correspondingly, quality thinning results in a value correction for the remaining trees as

$j = 1+b(1-p)$ (3),

where $+-b$ are the relative limits of the uniform distribution, set as $+-0.5$ throughout this paper, and $p$ is the survival rate of trees in the quality thinning. Eq. (3), if applied, is here applied to the first commercial thinning.

Quality thinning, however, cannot be applied on strip roads, where all the trees have to be removed. Then, after opening the strip roads, the survival rate of trees in the quality thinning is

$p=s/a$ (4),

where $s$ is the total survival rate in the harvesting, and $a$ is the survival rate after opening strip roads, but before any eventual quality thinning.

A somewhat nontrivial question is, how the value correction due to quality should be applied. Pulpwood-sized trees contain pulpwood only, regardless of their quality. However, good-quality pulpwood stems may later contain a larger proportion of sawlogs, or possibly high-value specialty logs. Let us write the expected value of the roadside price of a single trunk of specified size and species as

$$\langle P_{tree} \rangle = \langle P_{pulp} \rangle + \langle P_{saw} \rangle = \langle v_{pulp} \rangle \langle p_{pulp} \rangle + \langle v_{saw} \rangle \langle p_{saw} \rangle \qquad (5),$$

where the right-hand side quantities correspond to the expected volumes and unit prices. However, in the case of quality variation, most, if not all, of the quantities on the right-hand side vary. Introducing a statistical distribution into all these quantities would be a complicated task. An even more serious problem is that in the case of pulpwood size trees, the sawlog content is always zero, even if the tree quality obviously affects the sawlog value to be gained later.

Apparently, such problems can be avoided by assigning the quality variation into the sawlog unit price. Then, the roadside price of a trunk of specified size and species becomes

$$P_{tree} = \langle P_{pulp} \rangle + P_{saw} = \langle v_{pulp} \rangle \langle p_{pulp} \rangle + \langle v_{saw} \rangle p_{saw} = \langle v_{pulp} \rangle \langle p_{pulp} \rangle + \langle v_{saw} \rangle j \langle p_{saw} \rangle \qquad (6).$$

It is found from Eq. (6) that the quality only contributes to the roadside value if both the expected value of sawlog content and the expected value of sawlog price differ from zero, and then is proportional to both these quantities.

During further stand development after quality thinning, the correction coefficients given by Eq. (3) evolve according to Eq. (7):

$$j_{d,t+1} = \frac{NT_{d-1,t} j_{d-1,t} + NR_{d,t} j_{d,t}}{NT_{d-1,t} + NR_{d,t}} \qquad (7),$$

where $NT_{i,j}$ is the number of trees transferring from size class $i$ to size class $i+1$ at time step $j$, and $NR_{k,l}$ is the number of trees remaining in size class $k$ at time step $l$.

It is worth noting that trees belonging to smallest diameter class are the smallest trees where quality thinning possibly is applied. Correspondingly, the quality correction of trees transferring into the smallest diameter class is taken as unity.

As stated in the Introduction, a third procedure in the examination of the effects of quality thinning is the eventual inclusion of the tree vigor in the concept of quality. In the case of the inclusion, the correction coefficient $j$ also scales the tree diameter increment rate in the growth model.

*Example stands*

Measurement data from seven never-thinned Norway spruce (Picea abies) -dominated young stands are taken as the set of initial stand conditions. The stands, of age between 30 and 45 years, have been described in detail in earlier papers [Kärenlampi 2019b, 2020c,2021a, 2021b,2022b, 2022c]. Each stand was represented by a circular spot of ten meters of radius, within which the breast-height diameter of any tree was recorded. The commercial volume of any trunk was determined on the basis of earlier collected harvester data, as clarified in [Kärenlampi 2021b, 2022d].

Prices and expenses are here retained at the 2019 level, to retain comparability with earlier investigations [Kärenlampi 2019b, 2020c, 2021a, 2021b,2022b, 2022c]. Regeneration expenses are amortized first at the occurrence of final harvesting [Kärenlampi 2020a].

Technically, a time evolution from any initial condition is established according to the growth model [Bollandsås et al. 2008], in terms of 30-month timesteps, any field observation serving as an initial condition. The design of thinning procedures and rotation age maximizing the expected value of the rate of return on capital (Eq. (2)) was implemented as described in [Kärenlampi 2023, 2024], and not repeated here, to avoid plagiarism claims.

**Results**

The expected value of the return rate on capital without and with quality thinning is shown in Figs. 1a and 1b, respectively. It is found that quality thinning always improves the achievable return rate on capital, but the effect is small, appearing in the third significant digit of the return rate. One of the seven experimental stands shows a longer rotation with quality thinning. In both cases, two of the seven stands benefit from two thinnings, whereas five stands are thinned only once. In Figure 1b, quality thinning is economically applied only to the largest diameter classes.

The expected value of the return rate on capital with quality thinning, quality correlated with diameter increment rate, is shown in Fig. 1c. It is found that quality thinning, when quality correlates with growth rate, improves the achievable return rate on capital, the effect appearing in the second significant digit. Two of the seven experimental stands shows a shorter rotation than in the absence of quality thinning, two show a longer rotation. Only one of the seven stands benefits from two thinnings, six stands thinned only once. Unlike in Fig. 1b, quality thinning, contributing to growth rate, is applied in all sizes of trees remaining, however strongest with the largest remaining trees.

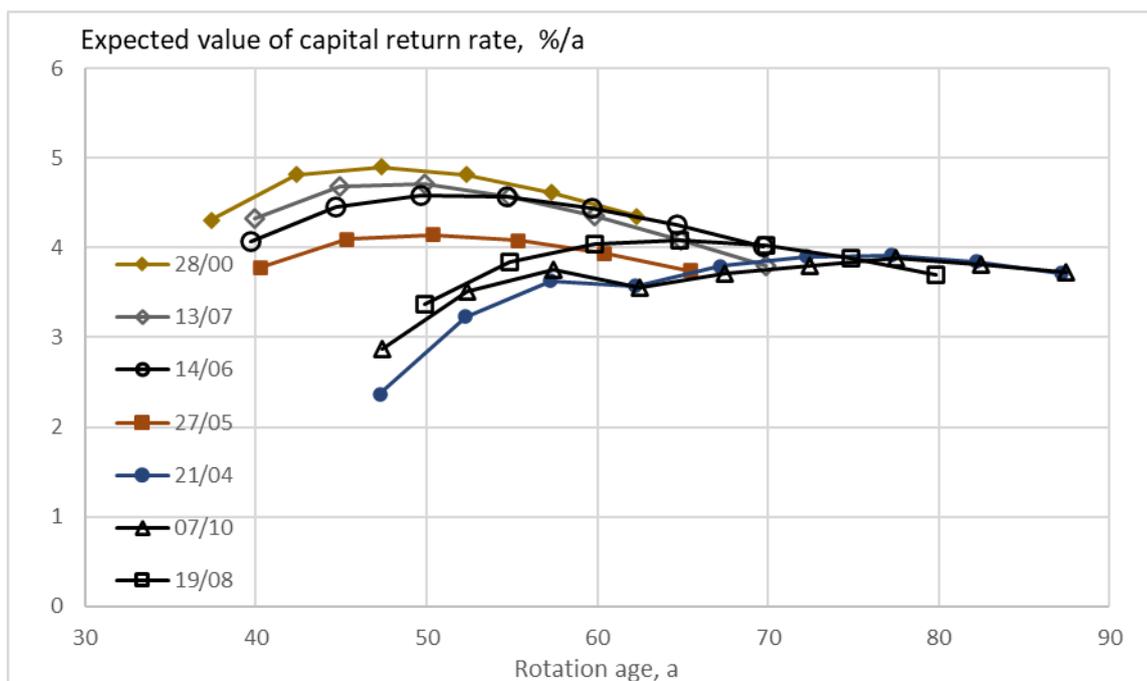

Fig. 1a. Expected value of return rate on capital in seven experimental stands, without quality thinning.

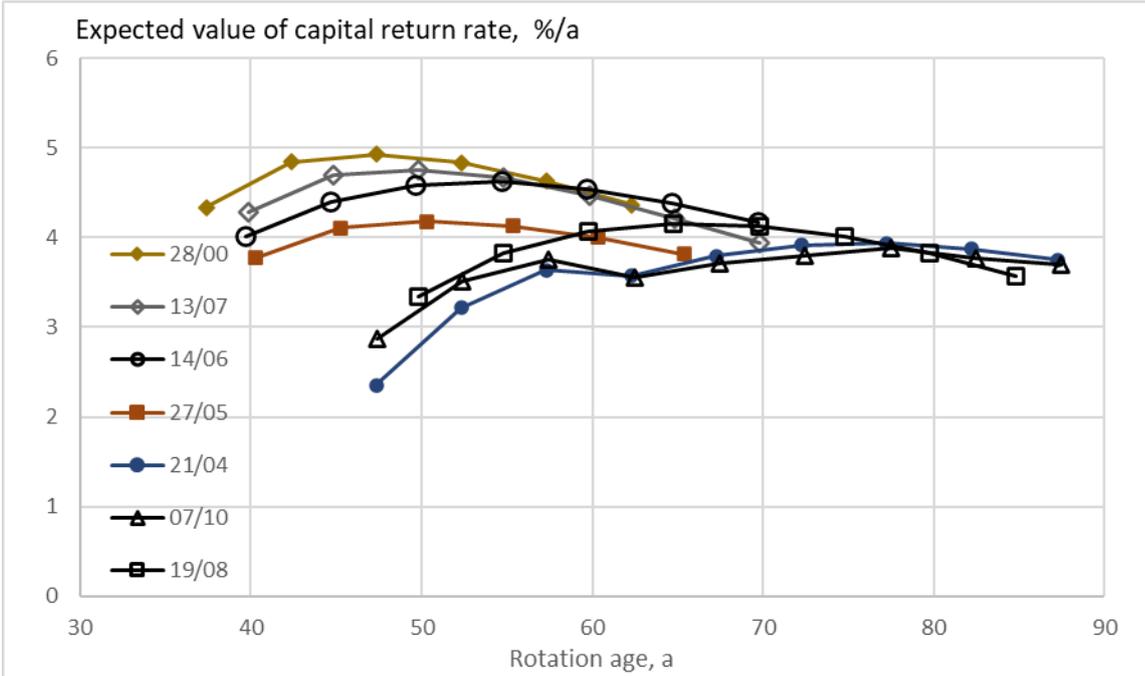

Fig. 1b. Expected value of return rate on capital in seven experimental stands, quality thinning applied.

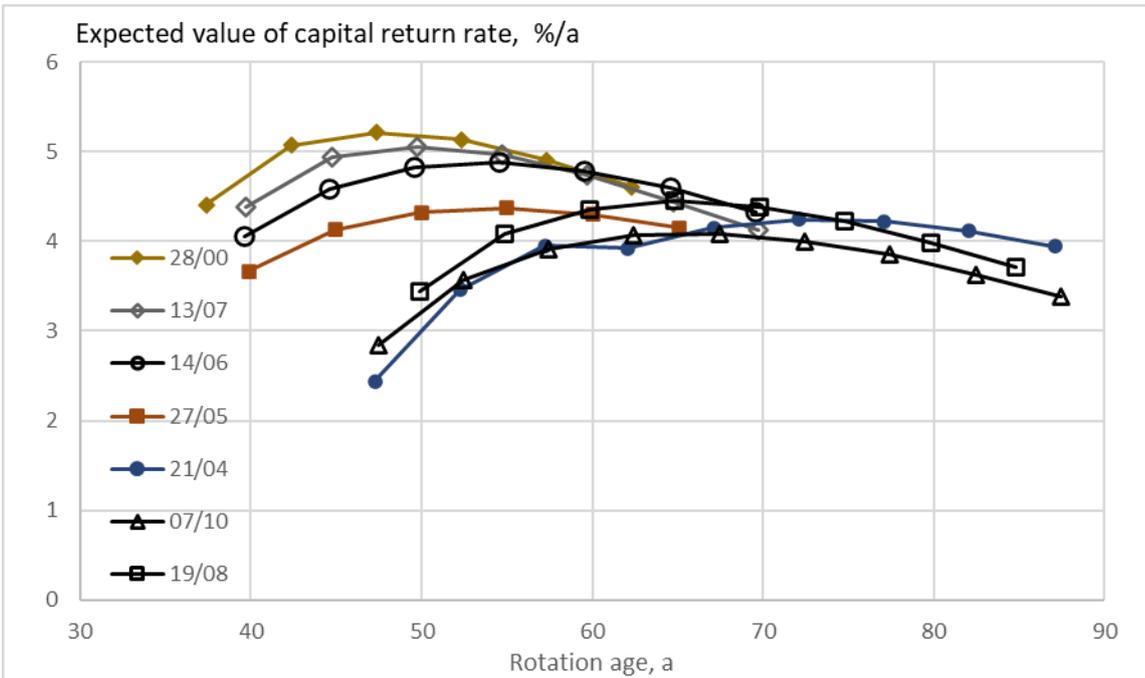

Fig. 1c. Expected value of return rate on capital in seven experimental stands, quality thinning applied, quality correlated with growth rate.

The basal area-weighted average breast-height diameter without and with quality thinning is shown in Figs. 2a and 2b, respectively. The data series in the Figures are terminated at the time of financial maturity. It is found that quality thinning does not affect the tree sizes much. Again, one of the seven experimental stands shows a longer rotation with quality thinning.

The basal area-weighted average breast-height diameter, quality correlated with diameter increment rate, is shown in Fig. 2c. It is found that quality thinning, when quality correlates with growth rate, induces a smaller decrement of average tree size in the first thinning, and a greater terminal tree size at stand maturity. Rotation ages exceeding 75 years are absent, but two stands show a longer rotation than without quality thinning.

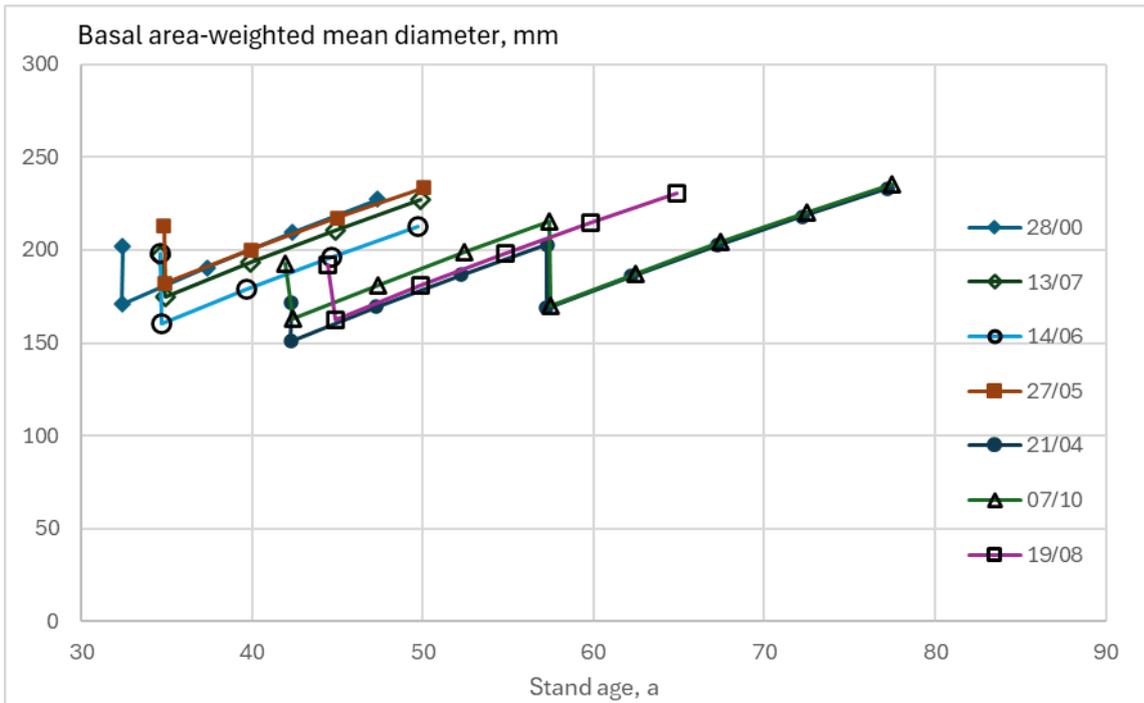

Fig. 2a. Basal area-weighted average diameter in seven experimental stands, without quality thinning.

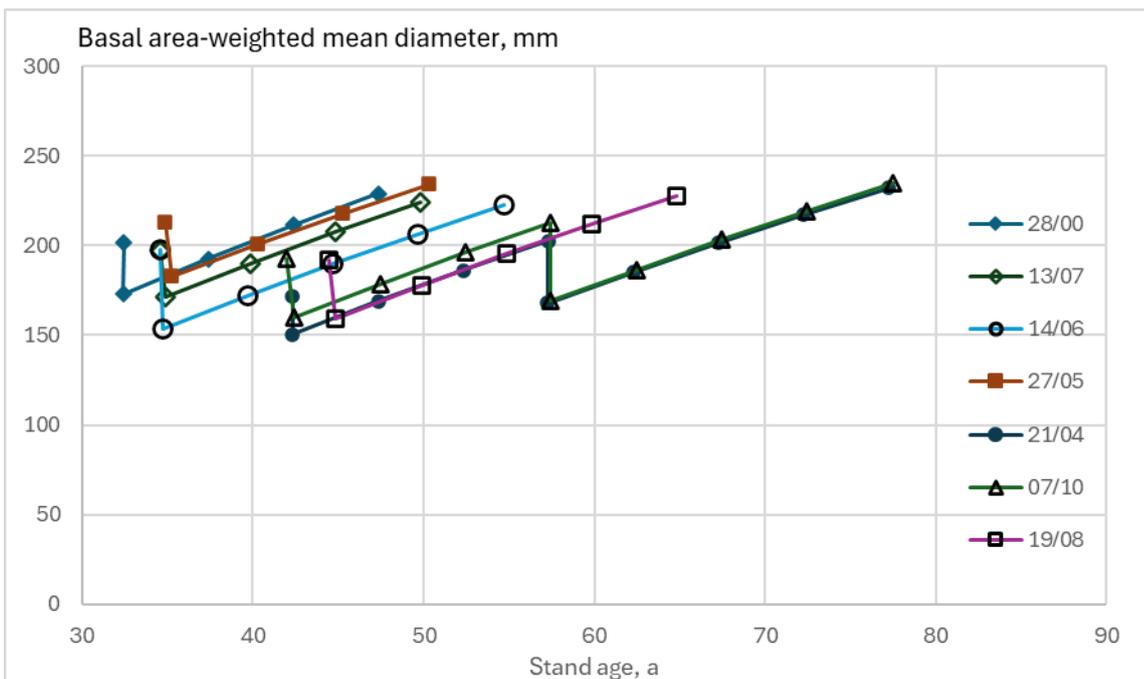

Fig. 2b. Basal area-weighted average diameter in seven experimental stands, quality thinning applied.

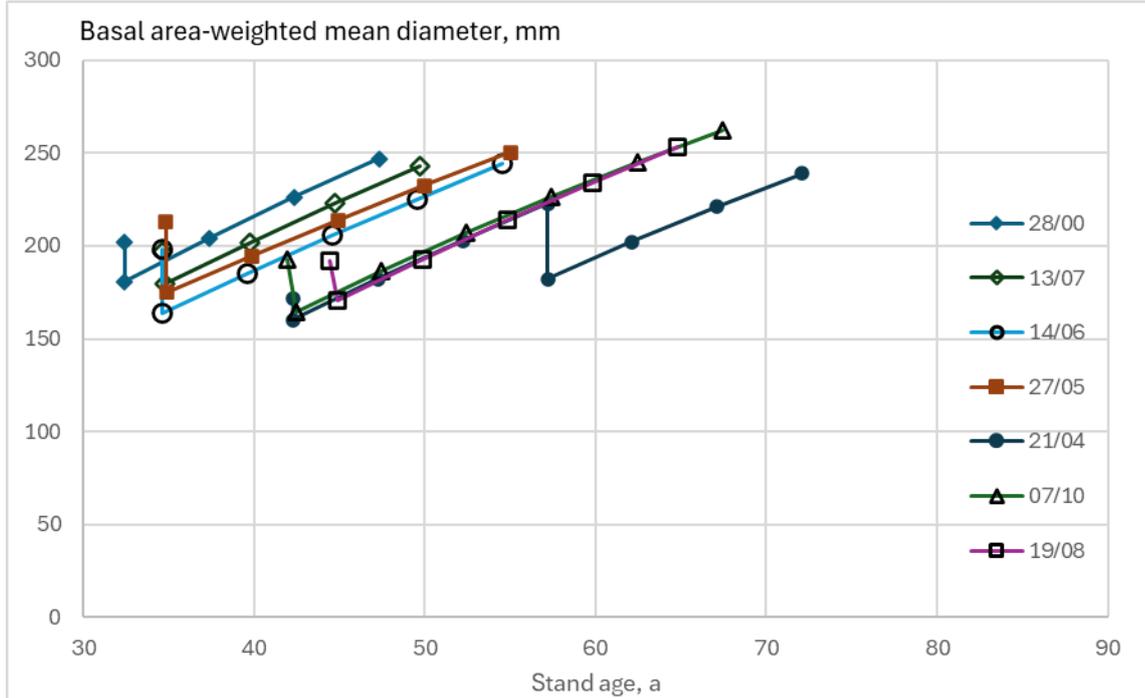

Fig. 2c. Basal area-weighted average diameter in seven experimental stands, quality thinning applied, quality correlated with growth rate.

Figure 3 shows the expected value of capitalization for the seven stands, without or with quality thinning, as a function of rotation age. It is found that the capitalizations do not vary much – there is more variation in the rotation age. The longest rotation ages refer to stands thinned twice (Fig. 1). When the quality correlates with growth rate, rotations are shorter, but the capitalization does not vary much.

The moderate variation in capitalization in Fig. 3 reveals an important result regarding the effect of eventual changes in bare land valuation on the quality thinning effect. A change in the bare land value $\Delta B$ contributes to the expected value of the return rate on capital as

$$\frac{\langle r \rangle^{`}}{\langle r \rangle} = \frac{\left\langle \dfrac{d\kappa}{dt} \right\rangle^{`}}{\left\langle \dfrac{d\kappa}{dt} \right\rangle} \frac{\langle K \rangle}{\langle K \rangle + \Delta B} \approx \frac{1}{1 + \dfrac{\Delta B}{\langle K \rangle}} \qquad (8),$$

where $\langle r \rangle^{`}$ is the changed value of the return rate on capital, and $\langle r \rangle$ is the original value. The latter equality in Eq. (8) is exactly correct if the expected value of the value increment rate (value growth rate) does not depend on the bare land value, or $\left\langle \dfrac{d\kappa}{dt} \right\rangle^{`} = \left\langle \dfrac{d\kappa}{dt} \right\rangle$. Since quality thinning does not affect much the expected value of capitalization $\langle K \rangle$ in Fig. 3, the contribution of the change in bare land value on the return rate on capital does not depend on quality thinning.

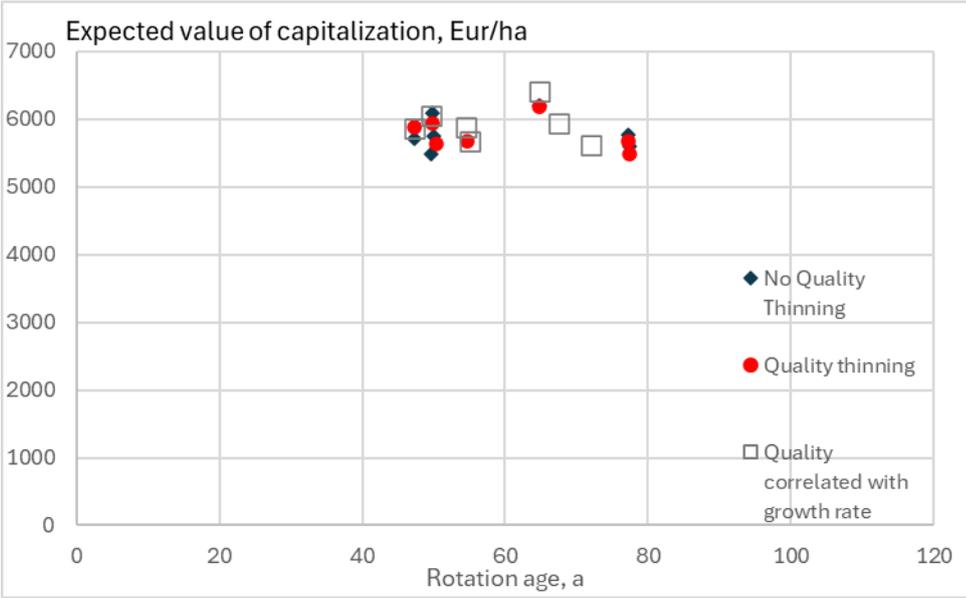

Fig. 3. Expected value of capitalization as a function of rotation age, with and without quality thinning.

Figure 4 shows the basal area-weighed mean value of diameter of the seven stands, without or with quality thinning, as a function of commercial stand volume at maturity. It is found that the quality thinning as such does not have much effect. The growth rate correlating with quality, trees grow bigger and the timber stock at maturity is somewhat reduced.

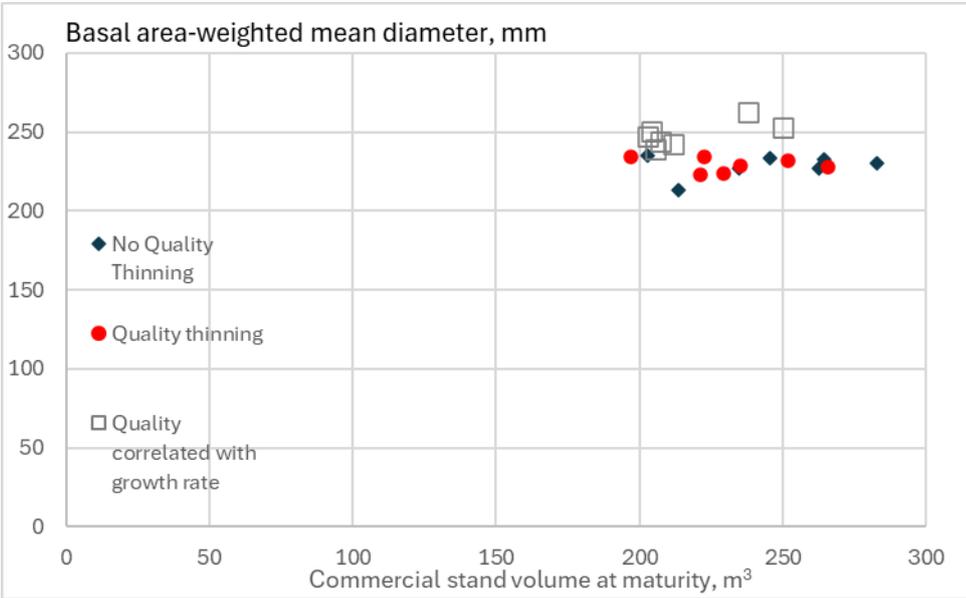

Fig. 4. Basal area–weighted mean diameter as a function of commercial stand volume at maturity, with and without quality thinning.

Differences appearing in Figs. 1 to 4 being due to different quality thinning procedures, it is of interest, how the quality parameter evolves in different tree size classes (according to Eq. (7)) after getting established according to Eq. (3). Without quality thinning, any quality parameter always has the value of unity. Figure 5 shows the expected values of quality parameters of spruce trees of different sizes at the event of financial maturity. It is found that in the case of quality thinning, without correlation to growth rate, the quality parameters are elevated only in the case of trees of at least 200 mm diameter. This is because quality thinning is applied only to the largest diameter classes remaining after thinning. In the case of Fig. 5b, quality correlating with growth rate, quality thinning is applied to all diameter classes, resulting in elevated quality parameters within a larger range of tree sizes.

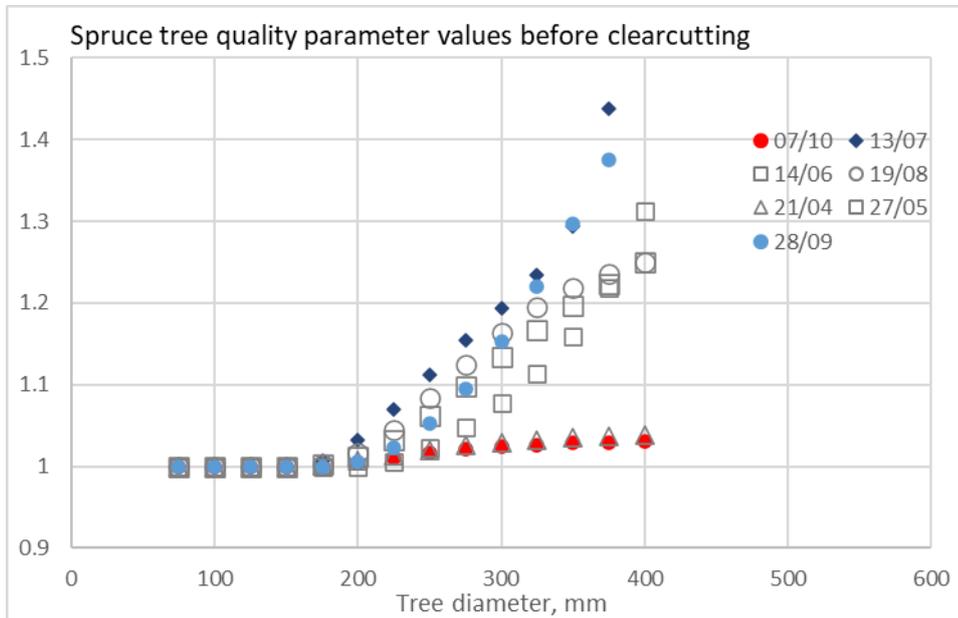

Fig. 5a. Expected values of quality parameters of spruce trees just before clearcutting, with quality thinning, quality not correlated to growth rate.

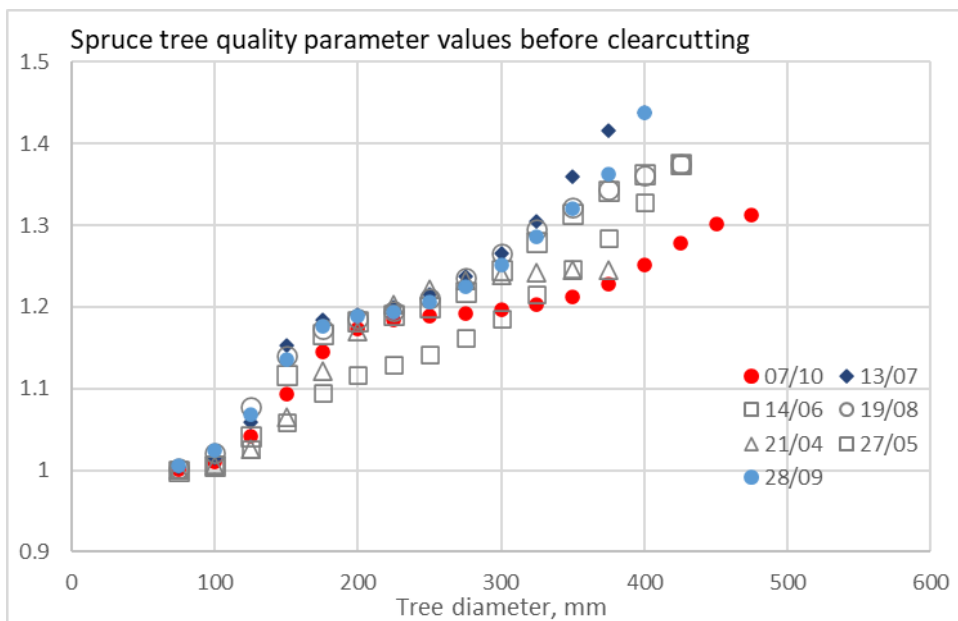

Fig. 5b. Expected values of quality parameters of spruce trees just before clearcutting, with quality thinning, quality correlated to growth rate.

In Fig. 6, the terminal quality parameters of birch trees differ from those of spruce trees, as smaller birch trees are retained in thinning, and even some small diameters of birch trees have been subjected to quality thinning. In the presence of the correlation with growth rate, quality thinning is more severe, resulting in larger quality parameters in Fig. 6b. It is again found from Fis. 5 and 6 that trees grow bigger in the presence of the correlation of quality to growth rate.

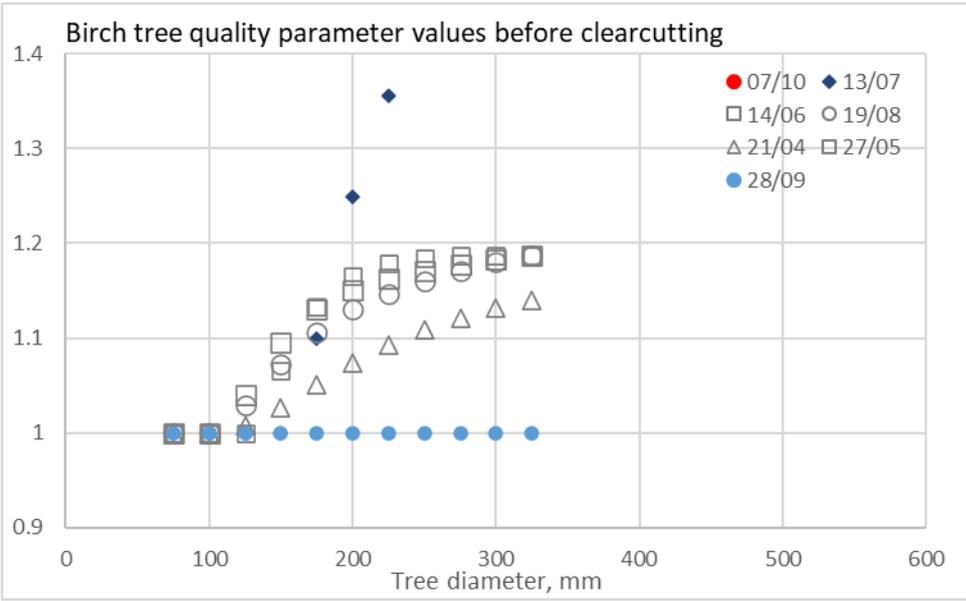

Fig. 6a. Expected values of quality parameters of birch trees just before clearcutting, with quality thinning, quality not correlated to growth rate.

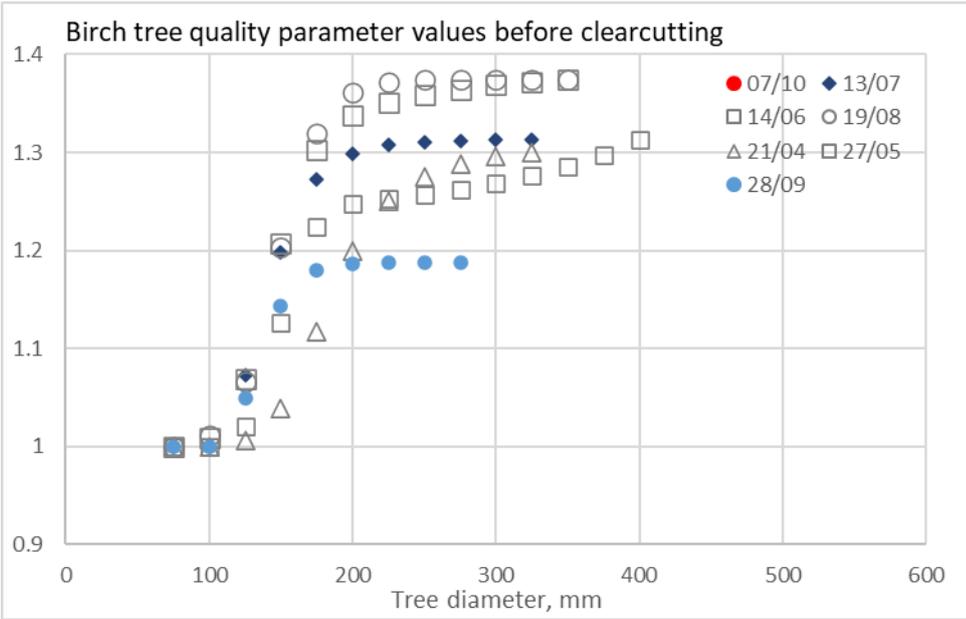

Fig. 6b. Expected values of quality parameters of birch trees just before clearcutting, with quality thinning, quality correlated to growth rate.

## Discussion

It is worth asking, what are the mechanisms affecting the relative value increment rate and correspondingly return rate on capital. Firstly, the steepness of value time gradients contributes. Secondly, phase transitions contribute to the steepness of the value gradients. Such effects can be investigated by plotting the relative value increment rate for different tree species and diameter classes.

The relative value development rate of different sizes of spruce and birch trees, respectively, at the time of stand observation, without commercial thinning, is shown in Figs 7 and 8. Both Figures show two separate peaks in the value increment rate. Firstly, trees in the smallest diameter class plotted increase strongly in roadside value since the harvesting expense per unit volume strongly reduces with increasing tree size. This effect is stronger in birch than in spruce, small birch trees being taller. Secondly, trees approaching the pulpwood-sawlog transition show another peak. This peak is sharper in the case of spruce trees, as the sawlog transition is sharper.

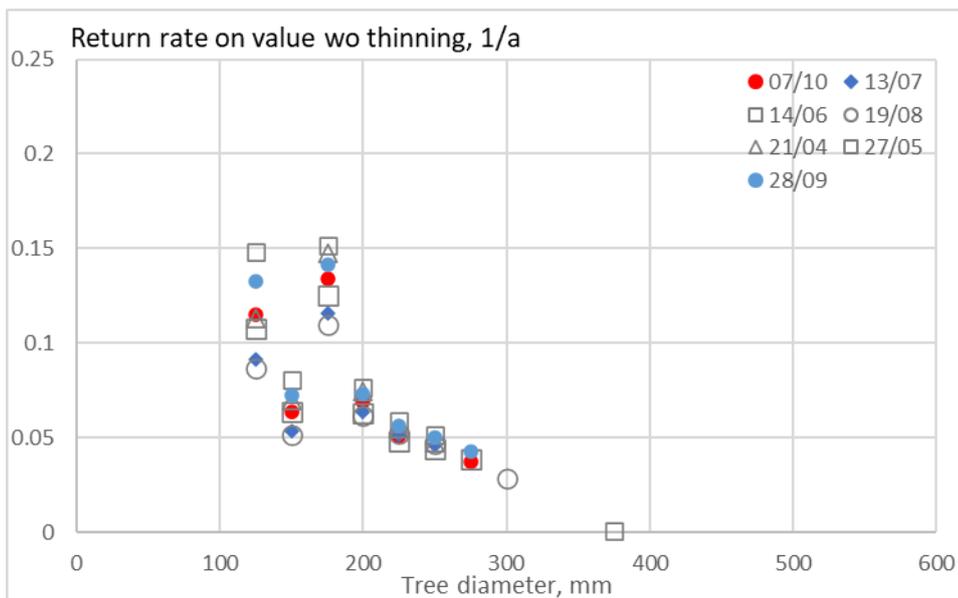

Fig. 7. Relative value increment rate of spruce trees, at the time of stand observation, without commercial thinning.

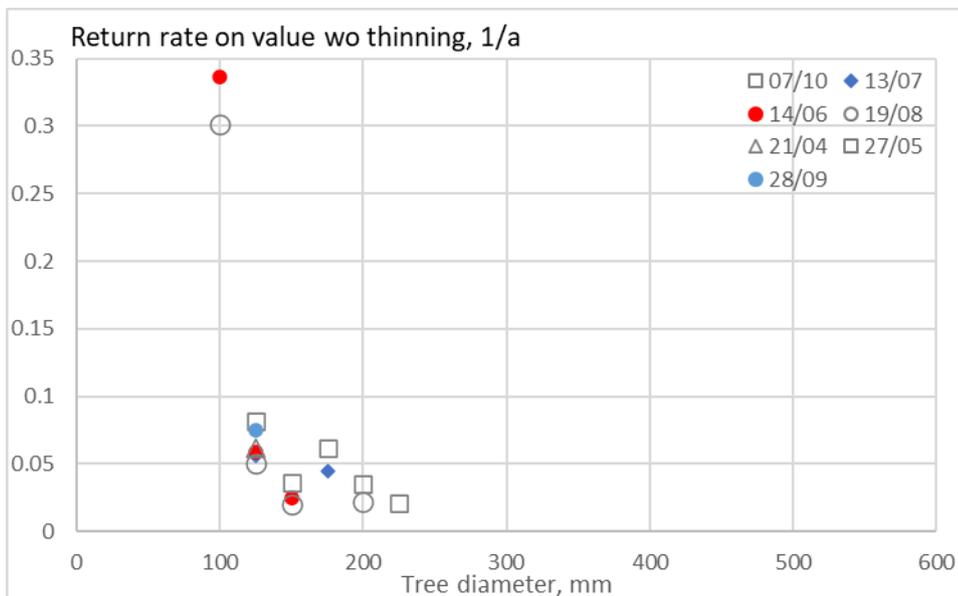

Fig. 8. Relative value increment rate of birch trees, at the time of stand observation, without commercial thinning.

The relative value development rate of different sizes of spruce trees, after non-quality thinning and quality thinning are shown in Figs. 9 and 10, respectively. A comparison to Fig. 7 shows that the value increment rate of small trees is greater, since thinning from above has opened growing space.

On the other hand, the biggest trees appearing in Fig. 7 are absent, as they have been removed in the thinning.

There are two observable differences between Figs. 9 and 10: in Fig. 10, the second peak of relative value increment rate is higher, corresponding to trees entering the pulpwood-sawlog transition. This obviously is an effect of the quality thinning applied to this size class. Secondly, Fig. 10 contains diameter classes that are absent from Fig. 9. Some high-quality trees have been retained in diameter classes exceeding 200 mm. However, this has happened on only two of the seven experimental stands. An obvious reason for the scarcity is that high-quality trees of that size already contain a larger proportion of valuable sawlogs. Correspondingly, there is less room for further relative value increment.

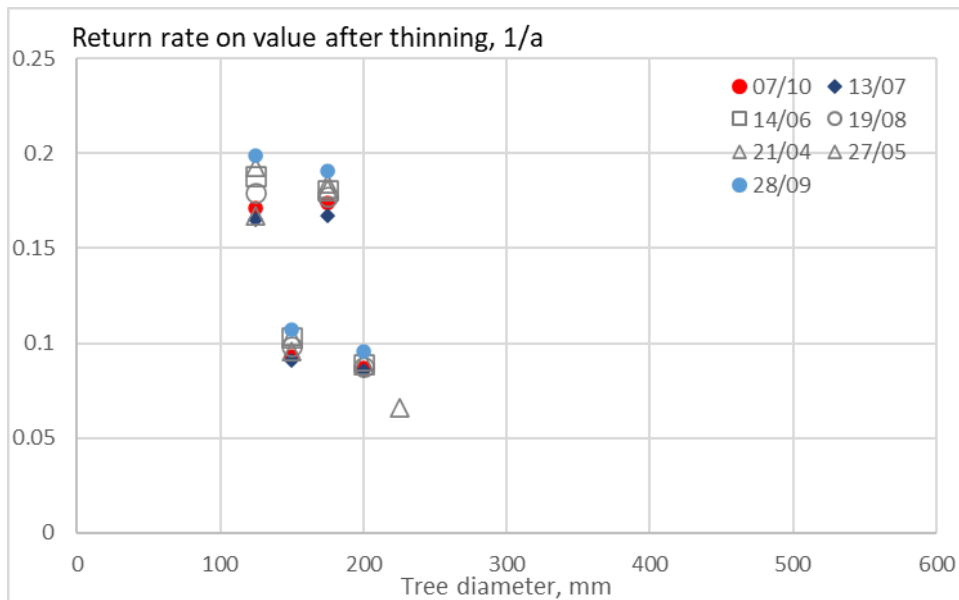

Fig. 9. Relative value increment rate of spruce trees, at the time of stand observation, after commercial non-quality thinning.

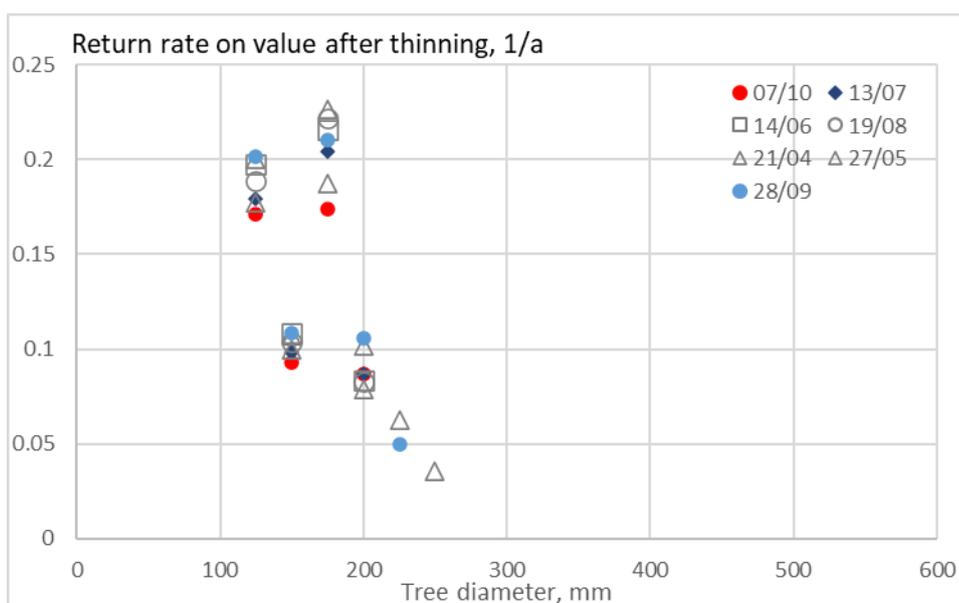

Fig. 10. Relative value increment rate of spruce trees, at the time of stand observation, after commercial quality thinning.

As Figures 9 and 10 show the relative value development rate after non-quality thinning or quality thinning, it is of interest to discuss eventual differences at later stages of stand development. Figure 11a shows the value development rate before clearcutting without quality thinning, and Figure 11b with quality thinning. The only obvious difference is the existence of bigger trees in the quality thinned case. This is because the quality thinning procedures have retained some trees in diameter classes that without quality thinning are removed in thinning from above.

It is worth noting that the relative value development rates of large trees in Figs. 11a and 11b are much larger than in Figs 7, 9 and 10. The reason is that the clearcutting premium of sawlogs [Kärenlampi 2022b, 2022a] is applied within the last stand development step.

Figure 11c shows the relative value increment rate before clearcutting, with quality thinning, quality correlated with growth rate. The prominent difference to Fig. 11b is the appearance of bigger trees, as the quality thinning has favored trees with the greatest growth rate. Another difference is that the level of value increment rates in Fig. 11c is greater in all diameter classes, as growth-rate-enhancing quality thinning has been applied in all diameter classes.

It is still worth noting that in the computation of the expected value of the return rate on capital, any momentary return rate is weighted by capitalization in Eq. (2). A similar weighting can be applied to different tree size classes in Figs. 9 and 10. Even if these two Figures differ, the contribution of the difference to the expected value of the return rate on capital within the rotation is rather small in Fig. 1, the value increment rate of small trees after thinning representing a small fraction of the capitalization accumulating during the rotation (the denominator of Eq. (2)).

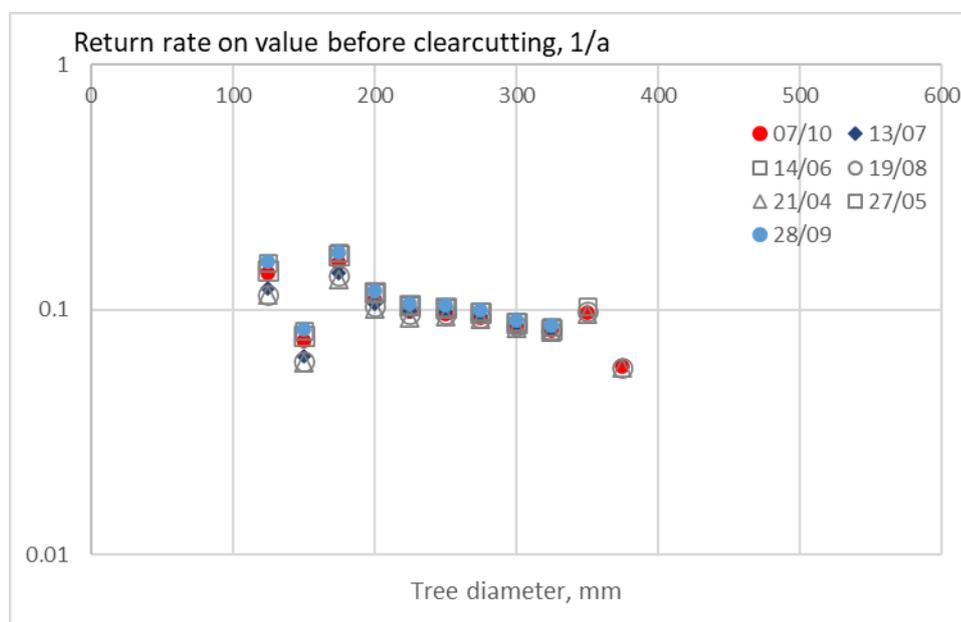

Fig. 11a. Relative value increment rate of spruce trees before clearcutting, without commercial quality thinning.

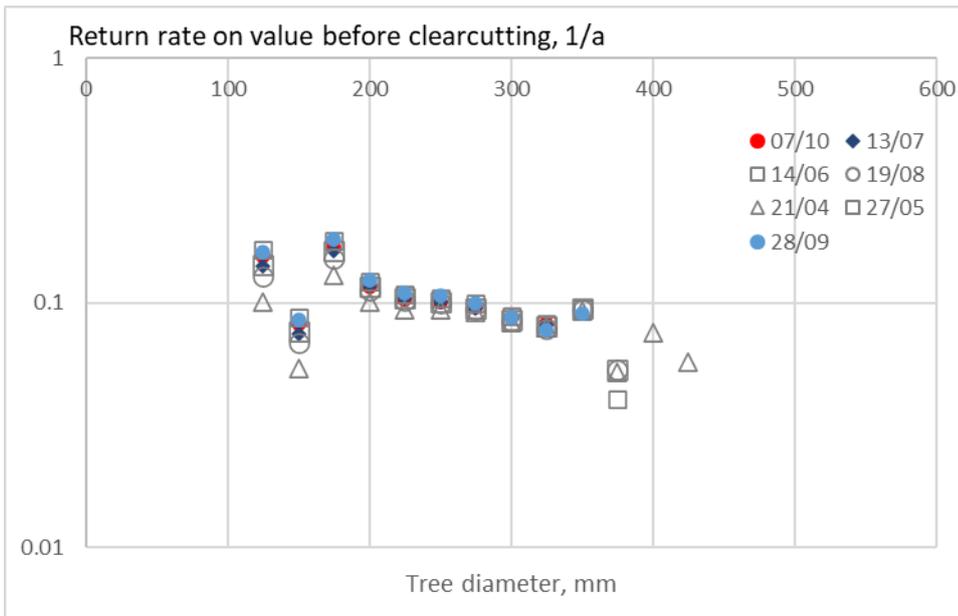

Fig. 11b. Relative value increment rate of spruce trees before clearcutting, after commercial quality thinning.

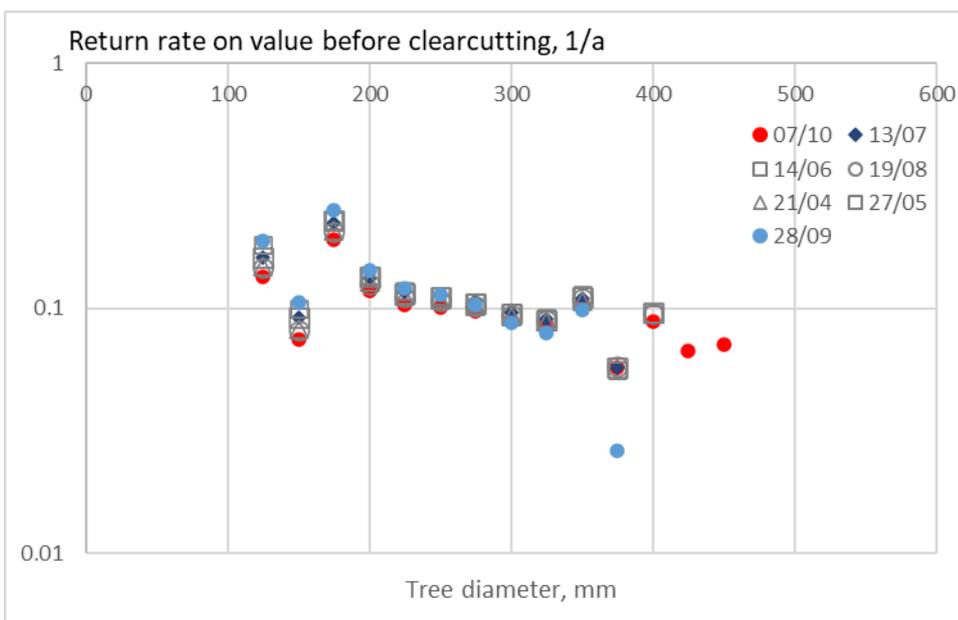

Fig. 11c. Relative value increment rate of spruce trees before clearcutting, after commercial quality thinning, quality correlated with growth rate.

In this paper, three procedures have been applied: non-quality thinning, quality thinning, and quality thinning with correlated growth rate. The assumption of a full correlation between quality and growth rate obviously is bold. However, some correlation may well appear, depending on how "quality" is interpreted by the harvester operator. In many cases, observable vigor might be included in the concept of quality [Cameron 2002], necessarily resulting in a correlation between quality and growth rate.

A somewhat surprising observation is the high relative value increment rate of the smallest trees in Figs. 7 to 11c. As no assortment transition takes place at that tree size range, the value increment rate is due to a considerable change of the harvesting expense per volume unit. The harvesting expense approximation is based on the productivity study of Nurminen et al. [2006]. The harvesting expense effect makes removal of small trees uneconomical, and thus leads to thinnings from above as the dominating strategy. This significant effect possibly deserves verification.

Empirical verification of the effect of stem size on the harvester time consumption per volume unit is shown in Fig. 12. There, results of Nurminen et al. [2006] are plotted for pine thinning and for birch clearcut, along with empirical time-consumption observations from a logging site of 4000 trunks from the summer of 2023, for two tree species. Additionally, the productivity curve computed from the harvesting price list of an anonymous entrepreneur is plotted. It is found that the empirical results reasonably verify the study of Nurminen et al. [2006]. The time consumption apparently follows a power law with an exponent -2/3. This indicates a power-law relationship between time consumption and diameter in the order of -2.

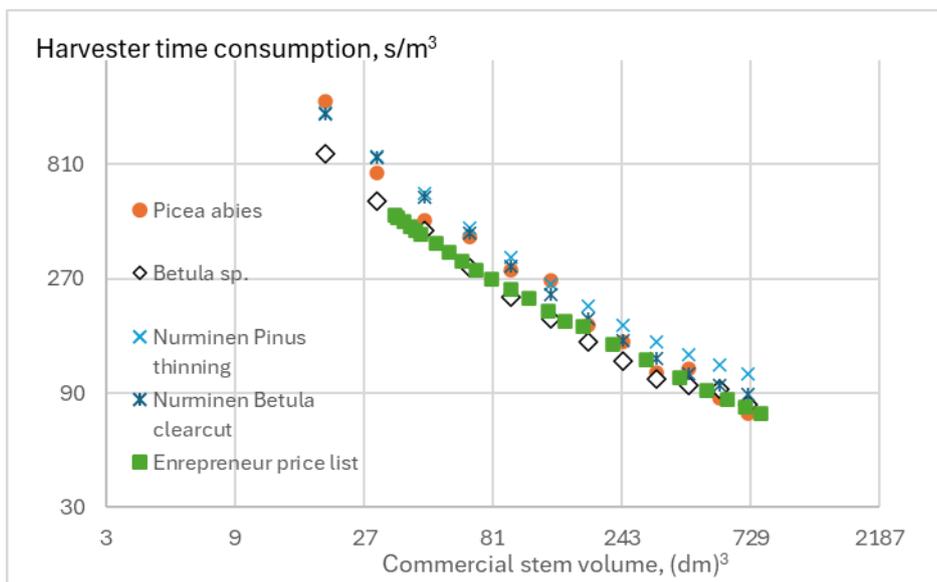

Fig. 12. Harvester time consumption per volume unit as a function of commercial stem volume, according to three different sources of information.

Quite a few studies have indicated a rather small effect of selective thinning of young stands on future sawlog yield [Niemistö et al. 2018, Nuutinen et al. 2021, Mielikäinen and Valkonen 1991, Resquin et al. 2024, Mäkinen et al. 2006, Karlsson et al. 2012, Segtowich et al. 2023]. Such results are essentially in line with the results of this paper. There are also contrasting views [Cameron 2002]. However, one must recognize that the financial considerations of this paper have not been implemented earlier in this context.

Heavy thinnings obviously enhance the risk of snow and wind damage [Cameron 2002, Cremer et al. 1982, Persson 1972, Valinger et al. 1993]. Most of the thinning procedures established in this paper are heavy. The economic quantification of the risk of snow and wind damage being difficult, the robustness of the treatment does not favor such a development. Heavy thinnings should be avoided on exposed sites. This can be practically implemented by replacing any heavy thinning with two gentle thinnings, a few years apart, giving the remaining trees time to adapt. Such an arrangement may not be adequate on seriously exposed sites [Cameron 2002].

It is worth asking what kinds of phenomena have not been considered in this study. Firstly, the present treatment is Markovian. Further development of any stand depended on the current state, regardless of history. Time delays exist, and further development depends on history.

Secondly, the quality distribution was described similarly for all trees, regardless of size and position within the stand. The expected yield of assortments was based on empirical data, considering tree size but not including any effect of the position of the tree within the stand. It is worth noting that the growth model does account for the position of a tree within the stand, in a Markovian manner.

The applied dataset consisted of never-thinned stands where thinning was due. There is a possibility that thinning on some of the stands was overdue. As there is a difference between Figs. 9 and 10 in the value increment rate of trees approaching the pulpwood-sawlog transition, the proportion of such trees certainly depends on the timing of the thinning.

## Acknowledgements

The author declares that no competing interests exist.
This work was partially funded by Niemi foundation. The funder had no role in study design, data collection and analysis, decision to publish, or preparation of the manuscript.